\documentclass[aps,epsfig,graphics,twocolumn,floatfix,mathbbm,a4paper,nofootinbib]{revtex4}

\usepackage{amsmath,amsfonts,amssymb,graphics,graphicx,epsfig,color,times,bbm}

\begin{document}
\bibliographystyle{apsrev}

\newcommand{\R}{\mathbbm{R}}
\newcommand{\rr}{\mathbbm{R}}
\newcommand{\E}{{\cal E}}
\newcommand{\cc}{{\cal{C}}}
\newcommand{\ii}{\mathbbm{1}}

\newcommand{\1}{\mathbbm{1}}
\newcommand{\F}{\mathbbm{F}}

\newcommand{\tr}[1]{{\rm tr}\left[#1\right]}
\newcommand{\gr}[1]{\boldsymbol{#1}}
\newcommand{\be}{\begin{equation}}
\newcommand{\ee}{\end{equation}}
\newcommand{\bea}{\begin{eqnarray}}
\newcommand{\eea}{\end{eqnarray}}
\newcommand{\ket}[1]{|#1\rangle}
\newcommand{\bra}[1]{\langle#1|}
\newcommand{\avr}[1]{\langle#1\rangle}
\newcommand{\D}{{\cal D}}
\newcommand{\eq}[1]{Eq.~(\ref{#1})}
\newcommand{\ineq}[1]{Ineq.~(\ref{#1})}
\newcommand{\sirsection}[1]{\section{\large \sf \textbf{#1}}}
\newcommand{\sirsubsection}[1]{\subsection{\normalsize \sf \textbf{#1}}}
\newcommand{\ack}{\subsection*{\normalsize \sf \textbf{Acknowledgements}}}
\newcommand{\front}[5]{\title{\sf \textbf{\Large #1}}
\author{#2 \vspace*{.4cm}\\
\footnotesize #3}
\date{\footnotesize \sf \begin{quote}
\hspace*{.2cm}#4 \end{quote} #5} \maketitle}
\newcommand{\eg}{\emph{e.g.}~}

\newcommand{\un}{1\mkern -4mu{\rm l}}
\newcommand{\proofend}{\hfill\fbox\\\medskip }


\newtheorem{theorem}{Theorem}
\newtheorem{proposition}{Proposition}

\newtheorem{lemma}{Lemma}

\newtheorem{definition}{Definition}
\newtheorem{corollary}{Corollary}

\newcommand{\proof}[1]{{\bf Proof.} #1 $\proofend$}

\newcommand{\alejo}[1]{{\color{red} #1}}

\newcommand{\uno}{1\!\!1}

\title{Large bipartite Bell violations with dichotomic measurements}

\author{C. Palazuelos$^1$ and Z. Yin$^{2,3}$}
\affiliation{
$^1$Instituto de Ciencias Matem\'aticas, Universidad Complutense de Madrid (Spain)\\
$^2$School of Mathematics and Statistics, Wuhan University, Wuhan (China)\\
$^3$Institute of Theoretical Physics and Astrophysics, University of Gda\'nsk, Gda\'nsk (Poland)}


\begin{abstract}
In this paper we introduce a simple and natural bipartite Bell scenario, by considering the correlations between two parties defined by general measurements in one party and dichotomic ones in the other. We show that unbounded Bell violations can be obtained in this context. Since such violations cannot occur when both parties use dichotomic measurements, our setting can be considered as the simplest one where this phenomenon can be observed. Our example is essentially optimal in terms of the outputs and the Hilbert space dimension.
\end{abstract}

\maketitle


The famous EPR's paper \cite{EPR1935} doubted the completeness of quantum theory and predicted that there should be a complete theory to explain nature which fulfilled locality and realism. We say that such a theory is classical or that it admits a local hidden variable model (LHVM). Whether nature can be explained by a LHVM became a philosophical rather than a physical debate for a long time. However, in his groundbreaking paper \cite{Bell}, Bell provided an inequality which must be satisfied by all classical probability distributions obtained in a certain measurement setting while, at the same time, is violated by some probabilities obtained from quantum measurements. This violation reveals that nonlocality (violation of local realism) is an intrinsic property of quantum theory, which plays an important role in quantum information science.

After Bell's work, a lot of Bell-type inequalities have been deeply studied (see \cite{BCPSW2014} for a review). The significance of studying violations of Bell inequalities is not only to quantify the deviation of classical and quantum theory, but it is also important in many other aspects of quantum information theory, such as entanglement witness \cite{HGBL2005,T2000}, quantum communication complexity \cite{BZPZ2004} and Hilbert dimension witness \cite{BPAGMS2008}. In the very last years, violations of Bell inequalities have become more popular because of their application to device independent quantum cryptography (\cite{A+2007}, \cite{AGM2006},  \cite{VaTh2}) and the generation of random numbers (\cite{Pironio+}, \cite{VaTh1}). It is natural that large violations of Bell inequalities provide more benefit, both in theory and applications. Moreover, large Bell violations have been also a suitable tool to study certain properties of quantum nonlocality and its relation with other resources (\cite{Buhrman et al. 2015}, \cite{CRS2012}, \cite{JP2011}, \cite{P2012}).

Remarkably, when we consider the probability distributions obtained by two observers using dichotomic measurements on a bipartite system, which is the simplest possible scenario, the ``amount of Bell violation" is upper bounded by a constant $1.676<K_G<1.783$, as it was proved by Tsirelson \cite{Tsirelson}. Note that in this case, we usually use expectation values to describe the correlations between the two parties, which is equivalent to use the joint probabilities. The corresponding inequalities are called correlation Bell inequalities. The limitation in this context leads to study two different generalizations: Multipartite correlation Bell inequalities and general bipartite Bell inequalities. Recent results have shown that in both contexts there exist unbounded Bell violations.

The research on unbounded violations of Bell inequalities has followed two parallel lines. One is based on operator space theory, which is prosperously developing in recent years as a branch of functional analysis. Through this approach, in \cite{PWJPV08} the authors showed that unbounded Bell violations exist for tripartite correlation Bell inequality. This answered in the negative a question posed by Tsirelson about the possibility of a similar result to the one existing in the bipartite case. In the general bipartite case, several works have shown that one can also obtain unbounded Bell violations and they have also studied how far these results are from being optimal (\cite{JP2011},\cite{JPPVW2010}). On the other hand, strong techniques from computer sciences have been also used in this problem. In this line, in \cite{BV2013}, \cite{BRSW2011} and \cite{Regev2012} the authors stablished and improved some of the above results in terms of nonlocal games.

Compare to the previous contexts, there is a simpler scenario, namely correlations can be defined by general measurements in one party and dichotomic measurements in the other. More precisely, suppose Alice can choose among $N$ different measurement settings labeled by $x=1,\ldots,N$. Each of them can result in one of $K$ outcomes, labeled by $a=1,\ldots, K$. Suppose also that Bob can choose among $N'$ different measurement settings labeled by $y=1,\ldots,N'$, but each of them has only binary outcomes, labeled by $b=\pm 1$. Then, if we denote by $P= (P(a,b|x,y))_{x,y}^{a,b}$ the associated probability distributions, we can define the following correlations between Alice and Bob:
\begin{equation*}
E(a|x,y) = P(a,1|x,y) - P(a,-1|x,y).
\end{equation*}

If the probability distribution $P= (P(a,b|x,y))_{x,y}^{a,b}$ admits a LHVM, then
\begin{equation*}
E(a|x,y) = \int_{\Omega} P(a|x,\lambda) E(y,\lambda) d\mathbb{P}(\lambda)
\end{equation*}
for every $a,x,y.$ Here, $(\Omega, \mathbb{P})$ is a probability space and $(P(a|x,\lambda))_a$ is a probability distribution for all $x,\lambda.$ Moreover $-1\leq E(y,\lambda)\leq 1$ for every $y,\lambda.$ We will denote the set of all LHV correlations by $\mathcal{L}.$

If $P$ has a quantum realization, then
\begin{equation*}
E(a|x,y) = tr (E_x^a \otimes E_y \rho)
\end{equation*}
for every $a,x,y$. Here, $\rho \in B(H_A \otimes H_B)$ is a quantum state, $(E_x^a)_x^a\subset B(H_A)$ are POVMs on Alice's part, i.e, $E_x^a \geq 0$ and $\sum_{a\in A} E_x^a =\un$ and $(E_y)_y \subset B(H_B)$ are dichotomic observables on Bob's part, i.e, $-\un\leq E_y \leq \un$ for every $y.$ We denote the set of all quantum correlations by $\mathcal{Q}.$

Following the standard definitions, for given natural numbers $N, N', K,$ we define an \emph{asymmetric Bell functional (inequality)} $M$ as a set $(M_{x,y}^{a})_{x,y}^a$ of real numbers with $x=1,\ldots, N$, $y=1,\ldots, N'$ and $a=1,\ldots, K$. For a given correlation $E= (E(a|x,y))_{x,y}^a,$ we define
\begin{equation}\label{eq:Bellineq}
\left\langle M,E\right\rangle  = \sum_{x=1}^{N}\sum_{y=1}^{N'}\sum_{a=1}^{K} M_{x,y}^{a} E (a|x,y).
\end{equation}

We define the classical and quantum bounds of $M$ respectively by
{\small
\begin{equation}\label{eq:bound}
B_{\mathcal{C}}=\sup\left\{ \left|\left\langle M, E\right\rangle\right|: \; E\in\mathcal{L}\right\},  \text{      }B_{\mathcal{Q}}=\sup\left\{ \left|\left\langle M,E\right\rangle\right|: \; E\in\mathcal{Q}\right\}.
\end{equation}}
Finally, the quantum violation of $M$ is defined by
\begin{equation*}
LV\left(M\right)=\frac{B_{\mathcal{Q}}\left(M\right)}{B_{\mathcal{C}}\left(M\right)}.
\end{equation*}

At this point, it is very natural to wonder whether one can get unbounded violations in this scenario, since it can be considered as the simplest context where such violations can occur. In this paper we will prove that there are indeed Bell inequalities as in (\ref{eq:Bellineq}) leading to unbounded violations. More precisely, we will show in the next section that for every $n$ there exists a Bell inequality $M$ with $N,N'=2^n$ and $K=n$ such that $LV\left(M\right)=\Omega(\sqrt{n}/(\log n)^2)$. Even though this example is still far from experimental realizations, the use of dichotomic measurements in Bob's party entails an important simplification with respect to the previous known examples of large violations. In addition, to obtain the previous order Alice and Bob can share the maximally entangled state and Alice can perform some simple von Neumann measurements on her party. Our approach is based on a modification of the Khot-Vishnoi (KV) game \cite{KV2005} and the results in \cite{BRSW2011}, to obtain an asymmetric version of it, which will work in our setting. Moreover, our result is near optimal (up to a logarithm factor) in terms of Alice's outputs and the Hilbert space dimension (see Appendix B). Although the motivation of our approach comes from Banach space theory, the results presented in this paper do not need any knowledge about it. In fact, we will present our results in terms of nonlocal games, where everything becomes very natural.

For a given nonlocal game $G=G(\pi, V)$, there are two natural ways to compare quantum and classical strategies. One is the quotient of the quantum and the classical value, denoted by $\omega_q(G)/\omega_c(G)$ (see Appendix A for a brief introduction about nonlocal games). Another interesting way is to consider the bias of the game. This is the probability of winning the game minus the probability of loosing the game. The classical bias $\beta(G)$ (resp. the quantum bias $\beta^*(G)$) is defined as the maximum bias over all possible classical strategies (resp. quantum strategies). We then consider the quantity $\beta^*(G)/\beta(G)$. As we will explain later (see Lemma 1 and Corollary 1), for any nonlocal game with binary answers in one party one has $\omega_q(G)/\omega_c(G)\leq 2$. Hence, there is no way to get large violations by looking at this quantity. However, we will give an example of these games for which the quantity $\beta^*(G)/\beta(G)$ is $\Omega(\sqrt{n}/(\log n)^2)$. To complete the picture, this is equivalent to find a Bell inequality of the form (\ref{eq:Bellineq}) for which the quantity $LV\left(M\right)$ is of the order above.
\section{Asymmetric Bell inequality with large violation}

In this section we will provide our main result. First, we will recall the KV game and then we will construct an ``asymmetric" version of it, which will be a nonlocal game with only two possible answers for Bob's questions. As we will show, the quotient between the quantum bias and the classical bias of the new game can be arbitrarily large.
\subsection{Khot-Vishnoi game}
For any $n = 2^l$ with $l \in \mathbb{N}$ and $\eta= 1/2-1/\log n$, we consider the group of all words in $\{0, 1\}^n$ and the Hadamard subgroup $H$ with $n$ Hadamard code words. The KV game \cite{BRSW2011} $G_{KV}$ is defined as follows: The referee chooses a uniformly random coset $[x]\in \{0,1\}^n/H$ and one element $z \in \{0, 1\}^n$ according to the probability distribution
$P(z_i = 1) = \eta, P( z_i = 0) = 1- \eta$ independently of $i$. Alice and Bob are asked questions $[x]$ and $[x\oplus z]$ respectively by the referee. They answer the outputs $a \in [x]$ and $b\in [x\oplus z]$, and they win the game if and only if $a \oplus b = z$. It is easy to see that the winning probability for a fixed strategy $P$ is:
\begin{equation*}
P_{win} = \frac{n}{2^n} \mathbb{E}_z \sum_{[x]} \sum_{a \in [x]} P (a, a\oplus z| [x], [x\oplus z]).
\end{equation*}

Notice that the number of possible inputs to each player is $2^n/n$ and the number of
possible outputs for each player is $n.$ By H. Buhrman, O. Regev, G. Scarpa and R. de Wolf's work \cite{BRSW2011}, we know the the violation of this game is
\begin{equation}\label{eq:violationforKV}
\frac{\omega_q(G_{KV})}{\omega_c (G_{KV})} \geq C \frac{n}{(\log n)^2},
\end{equation}since $\omega_c (G_{KV})\leq C_1\frac{1}{n}$ and $\omega_q(G_{KV})\geq C_2 \frac{1}{(\log n)^2}$. Here $C_1$, $C_2$ and $C$ are universal constants independent of the dimension. Interestingly, the previous value $\omega_q(G_{KV})$ is attained on the maximally entangled state in dimension $n$: $|\psi_n\rangle=1/\sqrt{n}\sum_{i=1}^n|ii\rangle$. More precisely, the corresponding quantum strategy is give by $\langle\psi_n|E_{[x]}^a\otimes E_{[y]}^b|\psi_n\rangle$, where for a given $c\in [w]$, $E_{[w]}^c=|u_c\rangle \langle u_c|$ for $u_c=1/\sqrt{n}\sum_{i=1}^n(-1)^{c(i)}|i\rangle$. In particular, it is known that the previous result is essentially optimal in the number of outputs and in the dimension of the Hilbert spaces (see \cite{JP2011}, \cite{P2014}).
\subsection{Asymmetric version of the KV game}
Let us start by fixing some notation. Given a coset $[x]\in \{0,1\}^n/H$, since $|[x]|=n$ we can identify (by means of a simple enumeration) the coset $[x]$ with the group $\{0,1\}^l$. Then, for a given element $a\in [x]$, we will denote by $\tilde{a}\in \{0,1\}^l$ its corresponding image. Moreover, for $\tilde{a},\tilde{b}\in \{0,1\}^l$ we denote $\langle \tilde{a}, \tilde{b}\rangle=\sum_{i=1}^l \tilde{a}_i\tilde{b}_i$. With this at hand, we can easily define the asymmetric version of the KV game $G_{KV}^{as}$: The referee chooses a uniformly random coset $[x]\in \{0,1\}^n/H$ and one element $z \in \{0, 1\}^n$ according to the probability distribution $P(z_i = 1) = \eta, P( z_i = 0) = 1- \eta$ independently of $i.$ Moreover, the referee chooses a uniformly random element $k\in \{0,1\}^l$. Then, Alice and Bob are asked questions indexed by $[x]$ and $([x \oplus z], k)$ respectively, and they will answer the outputs $a\in [x]$ and $b= \pm 1$. The players win the game if and only if $$(-1)^{\langle \tilde{a\oplus z}, k\rangle}=b,$$where $\tilde{a\oplus z}\in \{0,1\}^l$ is the element associated to $a\oplus z\in [x \oplus z]$.

Now, for a given strategy {\small $P= \big(P\big(a,b|[x],([y],k)\big)\big)_{[x],([y],k)}^{a,b}$}, it is straightforward  to check that $P_{win}-P_{loose}$ is equal to
{\small \begin{equation*}
\frac{1}{2^n} \mathbb{E}_z \sum_{[x]} \sum_{a\in[x]} \sum_{k\in \{0, 1\}^l} (-1)^{\langle \tilde{a\oplus z}, k\rangle} E\big(a|[x],([x\oplus z],k)\big),
\end{equation*}}where {\small $E\big(a|[x],([x\oplus z],k)\big)= P\big(a,1|[x],([y],k)\big)- P\big(a,-1|[x],([y],k)\big)$}. Hence, the classical bias (resp. quantum bias) of the game $G_{KV}^{as}$ can be understood as the classical bound (resp. quantum bound) of an asymmetric Bell inequality in the sense of (\ref{eq:Bellineq}).
\begin{theorem}\label{thm:asyKVgame}
If $G_{KV}^{as}$ denotes the asymmetric version of the Khot-Visnoi game introduced above, we have
\begin{equation}
\frac{\beta^*(G_{KV}^{as})}{\beta(G_{KV}^{as})}\geq C\frac{\sqrt{n}}{\log^2n},
\end{equation}
where $C$ is an universal constant.
\end{theorem}

The proof is based on the following estimates:
\begin{equation}\label{key inequalities}
\beta^*(G_{KV}^{as})\geq \omega_q(G_{KV}) \; \text{and} \; \beta(G_{KV}^{as})\leq \sqrt{n}\omega_c(G_{KV}).
\end{equation}Indeed, with these two inequalities at hand we can immediately obtained the statement by invoking Eq. (\ref{eq:violationforKV}). In order to see the first inequality in (\ref{key inequalities}), let us assume that Alice and Bob have a quantum strategy for the KV game defined by an entangled state  $|\psi\rangle$ and two families of POVMs $(E_{[x]}^a)_{[x]}^a$, $(F_{[y]}^b)_{[y]}^b$. Then, Alice and Bob can define another strategy for the asymmetric version of the KV game consisting of sharing the same quantum state and, moreover, Alice's strategy is also defined by the family $(E_{[x]}^a)_{[x]}^a$. On the other hand, for a given question $([y],k)$ to Bob, he will consider the self adjoint operator
\begin{equation*}
B_{[y],k}=\sum_{b\in [y]}(-1)^{\langle \tilde{b}, k\rangle}F_{[y]}^b.
\end{equation*}
It is clear that $-\un \leq B_{[y],k} \leq \un$ for every $([y],k)$. Hence, we conclude that $\beta^*(G_{KV}^{as})$ is lower bounded by
{\small \begin{align*}
&\frac{1}{2^n} \mathbb{E}_z \sum_{[x]} \sum_{a\in[x]} \sum_{k\in \{0, 1\}^l} (-1)^{\langle \tilde{a\oplus z}, k\rangle} \langle\psi|E_{[x]}^a\otimes B_{[x\otimes z],k}|\psi\rangle\\&=\frac{1}{2^n} \mathbb{E}_z \sum_{[x]} \sum_{\substack{a\in[x]\\ b\in [x\otimes z]}} \sum_{k\in \{0, 1\}^l} (-1)^{\langle a \tilde{\oplus} z \oplus \tilde{b}, k\rangle}\langle\psi|E_{[x]}^a\otimes F_{[x\otimes z]}^b|\psi\rangle\\&=\frac{n}{2^n} \mathbb{E}_z \sum_{[x]} \sum_{a\in[x]} \langle\psi|E_{[x]}^a\otimes F_{[x\otimes z]}^{a\oplus z}|\psi\rangle.
\end{align*}}
That is, for every strategy performed by Alice and Bob for the KV game, we can define another strategy for the asymmetric version of the KV such that the quantum bias for the second one is lower bounded by the quantum value for the first one.

To show the second inequality in (\ref{key inequalities}), let us fix a classical strategy (correlation) for the $G_{KV}^{as}$. It suffices to look at the extreme points, so we can assume that  $E\big(a|[x],([y],k)\big)=P(a|[x]) E([y],k)$ with $P(a|[x])\geq 0$ and $\sum_aP(a|x)=1$ for every $x,a$ and $-1\leq E([y],k)\leq 1$ for every $([y],k)$. Then, we define
{\small \begin{equation*}
Q(b | [y])= \sum_{k\in \{0, 1\}^l} (-1)^{\langle \tilde{b}, k\rangle}E([y],k) \text{   } \text{ for every  } \text{   }  [y], \text{   } b\in [y].
\end{equation*}}

We claim that
\begin{equation}\label{claim}
n^{-\frac{3}{2}}\sum_{b\in [y]}|Q(b| [y])|\leq 1  \text{     }  \text{     for every  } \text{     }  [y].
\end{equation}
Then, if we consider $P =n^{-\frac{3}{2}}\Big(P(a|[x]) Q(b|[y]) \Big)_{[x],[y], a, b},$ we can easily deduce that $\langle G_{KV}, P\rangle$ equals
{\small \begin{equation*}
\frac{1}{\sqrt{n}2^n}\mathbb{E}_z \sum_{[x]}\sum_{a\in[x]}\sum_{k\in \{0, 1\}^l} (-1)^{\langle \tilde{a\oplus z}, k\rangle}P(a|[x])E([x\oplus z],k).
\end{equation*}}
On the other hand, the fact that the KV game has positive coefficients (as a Bell inequality) guarantees that $\omega_c(G_{KV}) \geq \langle G_{KV}, P\rangle$, since we could always improve the previous value by modifying the $n^{-\frac{3}{2}} Q(b|[y])$'s so that all they are positive and they sum up to one.

Since the last expression above is the same as the classical bias of the asymmetric KV game when we consider the correlation  $\big(E \big(a|[x],([y],k)\big)\big)_{[x],([y],k)}^a$, we deduce our result.

Finally, in order to show our claim \eqref{claim}, we note that $\sum_{b\in [y]}|Q(b| [y])|$ is equal to
{\small \begin{align*}
&\sum_{b\in [y]}\Big| \sum_{k\in \{0, 1\}^l} (-1)^{\langle \tilde{b}, k\rangle}E([y],k)\Big| \\&\leq \sqrt{n}\Big(\sum_{b\in [y]}\Big| \sum_{k\in \{0, 1\}^l} (-1)^{\langle \tilde{b}, k\rangle}E([y],k)\Big|^2\Big)^{\frac{1}{2}}\\&=\sqrt{n}\Big(\sum_{b\in [y]}\sum_{k, k'\in \{0, 1\}^l} (-1)^{\langle \tilde{b}, k\oplus k'\rangle}E([y],k)E([y],k')\Big)^{\frac{1}{2}}\\&=n\Big(\sum_{k\in \{0, 1\}^l} E([y],k)^2\Big)^{\frac{1}{2}}\\&\leq n^{\frac{3}{2}}\sup_{k\in \{0, 1\}^l}|E([y],k)|\leq n^{\frac{3}{2}}.
\end{align*}}
Hence we complete the proof of Theorem \ref{thm:asyKVgame}.
\section{Bounded violation for any asymmetric Bell inequality with nonnegative coefficients}
The reader could find it surprising that, in order to get large Bell violations in our setting, we have considered the bias of the game $G^{as}_{KV}$ in Theorem \ref{thm:asyKVgame}, while in (\ref{eq:violationforKV}) the authors obtained large violations by looking at the values of $G_{KV}$. The reason is that our new context is much more restricted than the case of general answers in two parties and now we cannot expect the quotient $\omega_q(G^{as}_{KV})/\omega_c (G^{as}_{KV})$ to be large. This can also be understood as the fact that we cannot have large Bell violations of asymmetric Bell inequalities with nonnegative entries.
\begin{lemma}\label{lem:positive}
Let $M$ be a set $(M_{x,y}^{a})_{x,y}^a$ of nonnegative numbers. Then, with the notations in \eqref{eq:bound},
\begin{equation*}
B_{\mathcal{C}}(M)=B_{\mathcal{Q}}(M).
\end{equation*}
\end{lemma}

Indeed, note that for any family of POVMs for Alice $(E_x^a)_x^a\subset B(H_A)$ and dichotomic observable for Bob $(E_y)_y \subset B(H_B)$ we easily deduce that
\begin{equation*}
\big\|\sum_{x,y,a}M_{x,y}^{a} E_x^a\otimes E_y\big\|\leq \big\|\sum_{x,y,a}M_{x,y}^{a} E_x^a\big\|.
\end{equation*}
Hence, $B_{\mathcal{Q}}(M)= \big\|\sum_{x,y,a}M_{x,y}^{a} E_x^a\otimes E_y\big\|_{B(H_A\otimes H_B)}$ is upper bounded by
{\small \begin{align*}
\big\|\sum_{x,y,a}M_{x,y}^{a} E_x^a\big\|_{B(H_A)}= \sup_\rho \big|\sum_{x,y,a}M_{x,y}^{a} tr(\rho E_x^a)\big|\leq B_{\mathcal{C}}(M).
\end{align*}}Here, the last supremum runs over all states on $H_A$ and the last inequality is immediate since $\big(tr(\rho E_x^a)\big)_{x,a}$ is a family of classical probability distributions. This concludes the proof.

Hence, for any nonlocal game with binary answers for one player (e.g. for Bob), we have the following result.
\begin{corollary}\label{corollary positivity}
Let $G=G(\pi,V)$ be a nonlocal game with binary answers for one player. Then,
\begin{equation*}
 \omega_q(G) \leq 2 \omega_c(G).
\end{equation*}
\end{corollary}
For the proof we note that
{\small \begin{align*}
\omega_q(G) &= \sup \Big\| \sum_{x,y,a,b}\pi(x,y) V(a,b|x,y)  E_x^a\otimes F_y^b \Big\|_{B(H_A\otimes H_B)}\\
& \leq \sup\Big\|\sum_{x,y,a,b} \pi(x,y) V(a,b|x,y) E_x^a\otimes A_{(y,b)} \Big\|_{B(H_A\otimes H_B)}\\&=\sup \Big| \sum_{x,y,a,b} \pi(x,y) V(a,b|x,y) P(a|x) \alpha_{(y,b)} \Big|.
\end{align*}}Here, the first supremum runs over all pairs of POVMs $(E_x^a)_{x}^a, (F_y^b)_{y}^b$ for Alice and Bob respectively, while in the second one $-\un\leq A_{(y,b)} \leq \un$ for every $(y,b)$. The last equality follows Lemma \ref{lem:positive}, where now the last supremum runs over all classical strategies $(P(a|x))_{x}^a$ and $-1\leq \alpha_{(y,b)}\leq 1.$ Note that in order to use the lemma, we must view $M= (M_{x,(y,b)}^a)_{x,(y,b)}^a$ as an asymmetric Bell inequality, where $M_{x,(y,b)}^a = \pi(x,y)V(a,b|x,y).$ Finally, it is very easy to see that the last quantity is upper bounded by $2 \sup\big|\sum_{x,y,a,b} \pi(x,y) V(a,b|x,y) P(a|x)\alpha_y^b\big|$, where this supremum runs over all families $(\alpha_y^b)_{y}^b$ such that $|\alpha_y^0|+|\alpha_y^1|\leq 1$ for every $y$. Using that the numbers $\pi(x,y)V(a,b|x,y) P(a|x)$ are nonnegative one can easily conclude that the last supreumum is upper bounded by $\omega_c(G)$, and the proof follows.

Note that Corollary \ref{corollary positivity} can be stated in a more general way. Indeed, since we have not used the structure of nonlocal games, the same result can be stated for any nonnegative Bell inequality (acting on probability distributions) with dichotomic measurements in one party.
\section{Conclusions}
In this work we have provided an example of a bipartite Bell inequality with dichotomic measurements in one party, which can give unbounded Bell violations. Therefore, we can simplify the scenario for which one can obtain such violations. This simplification could be regarded as a new step towards possible experimental realizations. However, reducing the number of measurements seems to be a crucial point since, in our example, they scale exponentially. On the other hand, our result is essentially optimal in Alice's outputs and in the Hilbert space dimension. Additionally, the asymmetry considered in this work can also be interesting for device independent scenarios, where it seems very reasonable to assume that Alice and Bob have different kinds of measurements. Finally, from a computer science point of view our result can be considered as a new example of a nonlocal game, for which the quotient of the quantum bias and the classical bias can be arbitrarily large. Interestingly, we do need to consider the bias of the game since, in contrast with more general settings, in our context we cannot use the classical and the quantum value of the game to obtain large violations.

We would like to thank David P\'erez-Garc\'ia, Micha{\l} Horodecki, Marcin Marciniak and Marius Junge for many helpful discussions.

CP was supported by Spanish research projects MTM2011-26912, PRI-PIMCHI-2011-1071 (part of CHIST-ERA CQC), QUITEMAD+-CM, ref. S2013/ICE-2801 and ``Ram\'on y Cajal" program. ZY acknowledge support by TEAM project of FNP, ERC AdG grant QOLAPS and NSFC under Grant No.11301401.
\section*{Appendix A: Nonlocal games}
A nonlocal game $G=G(\pi, V)$ is defined as follows \cite{CHTW2004}: $\pi$ is a probability distribution on $X \times Y,$ and $V:X \times Y \times A \times B\rightarrow \{0,1\}$ is the predicate function, where $X, Y, A, B$ are nonempty finite sets. The referee will randomly choose questions $(x,y) \in X \times Y$ according to $\pi,$ and send the questions to two players Alice and Bob. The players (without communication) will answer the questions by $(a,b) \in A \times B.$ They win the game if and only if $V(a,b|x,y)=1.$ Before the game starts, Alice and Bob may agree on some strategy (classical or quantum) to play the game.

The classical value of a game $G$ is the maximal winning probability, which is restricted by only using classical strategies. Thus, we have
\begin{equation*}
\omega_c(G) = \max_{a,b} \sum_{x,y} \pi(x,y) V(a(x),b(y)|x,y),
\end{equation*}
where the maximum is taken over all functions $a : X\rightarrow A$ and $b : Y \rightarrow B.$

If there is a quantum strategy for the players, i.e, there is a quantum state $\rho \in B(H_A \otimes H_B)$ shared by Alice and Bob, and a quantum measurement for Alice (Bob) for each $x\in X$ ($y\in Y$). For every question $(x,y),$ the probability to output $(a,b)$ is given by
\begin{equation*}
P(a,b|x,y) = tr (E_x^a \otimes E_y^b \rho),
\end{equation*}
where $(E_x^a)_x^a\subset B(H_A)$ and $(E_y^b)_y^b \subset B(H_B)$ are POVMs for Alice and Bob respectively. Thus, the quantum value of the game is

\begin{equation*}
\omega_q(G) = \sup \sum_{x,y} \sum_{a,b} \pi(x,y) V(a,b|x,y) tr (E_x^a \otimes E_y^b \rho),
\end{equation*}
where the sup is taken over all possible quantum strategies. We define the violation of the game $G$ as the ratio of quantum and classical value, i.e, $\omega_q(G)/\omega_c(G).$
\section*{Appendix B:  Near optimal violation for asymmetric Bell inequalities}
An interesting property of the KV game is that it offers a near optimal violation in terms of the number of outputs and the dimension of the Hilbert space. Indeed, for any Bell inequality with $N$ inputs, $K$ outputs and Hilbert space dimension $d$, we know that the optimal violation is $O(m)$, with  $m=\min\{K,N,d\}$ \cite{JP2011}. In this section, we will provide sharp upper bounds for any asymmetric Bell inequality in terms of the output (of Alice) and the Hilbert space dimension of Bob. As a corollary of the propositions below, the asymmetric KV game offers an explicit example of an asymmetric Bell inequality whose violation is essentially optimal (up to a logarithm factor) in the number of Alice's outputs and Hilbert space dimension of Bob.
\subsection{Upper bounds in terms of the number of Alice's outputs}
\begin{proposition}\label{prop:nearoptimal}
For any asymmetric Bell inequality $M$ as in \eqref{eq:Bellineq}, the largest violation is $O(\sqrt{K-1}).$
\end{proposition}
The proof of this proposition is based on Grothendieck's inequality for complex matrices and it follows the same ideas as in (\cite[Section 5]{JP2011}). Given a complex matrix $(M_{x,y})_{x,y}$ with $x=1,\cdots, N$, $y=1,\cdots, N'$, let us denote
{\small \begin{align*}
&B_{\mathcal{C}}(M)=\sup\big\{\big|\sum_{x,y}M_{x,y}t_x s_y\big|\big\} \text{    } \text{  and   } \\&B_{\mathcal{Q}}(M)=\sup\big\{\big|\sum_{x,y}M_{x,y}\langle u_x, v_y\rangle \big|\big\},
\end{align*}}where the first supremum runs over all complex numbers $t_x$, $s_y$ with $|t_x|,|s_y|\leq 1$ for all $x, y$ and the second supremum runs over all complex Hilbert spaces $H$ and vectors $u_x$, $v_y$ in the unit ball of $H$ for all $x, y$.  Grothendieck's inequality (see for instance \cite[Ch.14]{DeFl1993}) states that there exists a positive constant $C$ such that for any complex matrix $(M_{x,y})_{x,y}$
\begin{equation}\label{Grothendieck ineq}
B_{\mathcal{Q}}(M)\leq C B_{\mathcal{C}}(M).
\end{equation}The smallest constant verifying inequality (\ref{Grothendieck ineq}) is called \emph{(complex) Grothendieck's constant} $K_G^{\mathbb C}$ and, although its exact value is still unknown, it verifies $1.338 \leq K_G^{\mathbb C}\leq 1.405$.

For a family of complex numbers $E= (E(x,y))_{x,y},$ we will denote
\begin{equation*}
\left\langle M,E\right\rangle  = \sum_{x=1}^{N}\sum_{y=1}^{N'} M_{x,y} E (x,y).
\end{equation*}
For the proof of the proposition it suffices to consider the case $N=N',$ i.e, Alice and Bob have same number of inputs. Let us consider an arbitrary asymmetric Bell inequality $M=(M_{x,y}^a)_{x,y}^a$ with $x,y=1,\ldots, N$ and $a=1,\ldots ,K$. Then, we define the following $(N+1)(K-1)\times N$ complex matrix
\begin{equation*}
\begin{cases}
M'_{(x,s),y}  = \sum_{a=1}^{K-1} \omega^{as} (M_{x,y}^a-M_{x,y}^K), &  \substack{x,y=1,\ldots, N\\s=1,\ldots, K-1}, \cr
M'_{(N+1,s),y} = \sum_{x=1}^N \omega^s M_{x,y}^K , &  \substack{y=1,\ldots,N\\s=1,\ldots, K-1},
\end{cases}
\end{equation*}
where $\omega = \exp{\frac{2\pi i}{K-1}}.$ According to (\ref{Grothendieck ineq}) we have that
\begin{equation*}
B_{\mathcal{Q}}(M') \leq K_G^{\mathbb{C}} B_{\mathcal{C}}(M').
\end{equation*}

Our upper bound will follow from the estimates
{\small \begin{equation}\label{Estimates upper bound K}
B_{\mathcal{C}}(M')  \leq 24(K-1)^{\frac{3}{2}} B_{\mathcal{C}}(M)\;\; \text{and} \;\; B_{\mathcal{Q}}(M') \geq (K-1)B_{\mathcal{Q}}(M).
\end{equation}}.

Let us start proving the second inequality. To this end, let us consider an arbitrary quantum strategy $\big(tr(\rho E_x^a\otimes B_y)\big)_{x,y=1;a=1}^{N;K}$ for $M$. In fact, by convexity we can assume that $\rho$ is a pure state, so that our strategy is of the form $\big(\big\langle \psi|E_x^a\otimes B_y|\psi\rangle\big)_{x,y=1;a=1}^{N;K}$, where $|\psi\rangle$ is a unit vector. Then, we define the following family of operators
\begin{equation*}
\begin{cases}  A_{(x,s)}=\sum_{a=1}^{K-1} \omega^{-as} E_x^a,  &   \substack{x=1,\ldots, N\\s=1,\ldots, K-1}, \cr
   A_{(N+1,s)}= \omega^{-s} \un,  & s =1,\ldots,K-1 \, .
   \end{cases}
 \end{equation*}

It is straightforward to check that $\|A_{(x,s)}\|\leq 1$ for every $(x,s)$. Therefore, $u_{x,s}=(A_{(x,s)}\otimes \un)|\psi\rangle$ and $v_{y}=(\un \otimes B_y)|\psi\rangle$ form a family of vectors in the unit Ball of a complex Hilbert space such that $\gamma=\big(\langle u_{x,s}, v_y\rangle\big)_{(x,s),y}=\big(\big\langle \psi|A _{(x,s)}\otimes B_y|\psi \big \rangle\big)_{(x,s),y}$. Thus, $B_{\mathcal{Q}}(M')$ is lower bounded by
{\small \begin{align*}
|\langle M', \gamma\rangle|&=\Big|\sum_{x=1}^{N+1}\sum_{y=1}^{N}\sum_{s=1}^{K-1} M'_{(x,s),y}\big\langle \psi|A _{(x,s)}\otimes B_y|\psi \big \rangle\Big|\\&
 = \Big| \sum_{x,y=1}^N\sum_{s,a,a'} \omega^{(a-a')s}(M_{x,y}^a-M_{x,y}^K) \big\langle \psi|E_x^{a'}\otimes E_y|\psi \big \rangle \\&+ \sum_{x,y=1}^N \sum_{s=1}^{K-1} M_{x,y}^K \big\langle \psi|\un\otimes E_y|\psi \big \rangle \Big|\\&
 = (K-1)\Big|\sum_{x,y=1}^N \sum_{a=1}^K M_{x,y}^a \big\langle \psi| E_x^a\otimes E_y|\psi \big \rangle\Big|.
\end{align*}}
For the last equality we have used that $\sum_{a=1}^{K-1}E_x^a= \un- E_x^K$ for every $x=1,\ldots, N$. This proves the second estimate in (\ref{Estimates upper bound K}).

In order to see the first estimate in (\ref{Estimates upper bound K}), let us consider an element  $(R(x,s)\beta_y)_{x,s,y},$ such that $|R(x,s)|\leq 1$ and $|\beta_y|\leq 1$ for all $x=1,\ldots, N+1, s=1,\ldots,K-1, y=1,\ldots,N$. Then, we define the following object $P$: For every $x=1,\cdots, N;$
\begin{equation*}
\begin{cases} P(a|x)=\sum_{s=1}^{K-1}  \omega^{as}  R(x,s), \;\; a=1,\ldots ,K-1, \cr
P(K|x)=\sum_{s=1}^{K-1} \omega^{s}  R(N+1,s)-\sum_{a,s=1}^{K-1} \omega^{as} R(x,s).
\end{cases}
\end{equation*}

\

Note that, for fixed $x=1,\ldots, N$,
{\small \begin{align*}
 \sum_{a=1}^K P(a|x)&= \sum_{a,s=1}^{K-1} \omega^{as} R(x,s)+\sum_{s=1}^{K-1} \omega^{s} R(N+1,s)\\&-\sum_{a,s=1}^{K-1}\omega^{as} R(x,s)=\sum_{s=1}^{K-1} \omega^{s} R(N+1,s),
 \end{align*}}
 which is a constant independent of $x$.

On the other hand,
{\small \begin{equation*}
\sum_{a=1}^K|P(a|x)| \leq 2 \sum_{a=1}^{K-1}\Big|\sum_{s=1}^{K-1}\omega^{as} R(x,s) \Big| + (K-1)\leq 3(K-1)^{3/2},
\end{equation*}}where the last inequality is proved exactly in the similar way as claim \eqref{claim} in Theorem \ref{thm:asyKVgame}.

The previous two properties joint with \cite[Lemma 3.2]{P2014} guarantee the existence of two classical strategies $P_1$ and $P_2$ for Alice and $\lambda_1,\lambda_2 \in \mathbb R$ such that $\Re(P)=\lambda_1 P_1+\lambda_2 P_2$ and $|\lambda_1|+|\lambda_2|\leq 3(K-1)^{3/2}$. Here, $\Re(P)$ denotes the real part of $P$. The same argument holds for the imaginary part $\Im(P)$, i.e, there exist classical strategies $P_3$ and $P_4$ for Bob and $\lambda_3,\lambda_4 \in \mathbb R$ such that $\Im(P)=\lambda_3 P_3+\lambda_4 P_4$ and $|\lambda_3|+|\lambda_4|\leq 3(K-1)^{3/2}$. On the other hand, for every $y$ we can write $\beta_y=\beta_y^1+ i \beta_y^2$ with $\beta_y^1$,  $\beta_y^2$ real numbers verifying $|\beta_y^j|\leq 1$ for $j=1,2$. Hence, our estimate will follow from the fact that $24(K-1)^{3/2}  B_{\mathcal{C}}(M)  $ is lower bounded by
{\small \begin{align*}
&3 (K-1)^{3/2}\sum_{i=1}^4 \sum_{j=1}^2\Big|\big\langle M, (P_i(a|x)\beta_y^j)_{x,y,a}\big\rangle\Big| \\&\geq \Big|\langle M, ( P(a|x)\beta_y)_{x,y,a}\rangle\Big| =\Big|\sum_{x,y=1}^N\sum_{a,s=1}^{K-1}M_{x,y}^a \omega^{as} R(x,s)\beta_y \\&+ \sum_{x,y=1}^N M_{x,y}^K \Big[\sum_{s=1}^{K-1}\omega^{s} R(N+1,s)-\sum_{a,s=1}^{K-1}\omega^{as} R(x,s)\Big]\beta_y\Big|\\&= \Big|\sum_{x,y=1}^N\sum_{a,s=1}^{K-1}(M_{x,y}^a-M_{x,y}^K)\omega^{as} R(x,s)\beta_y \\&+ \sum_{x,y=1}^N M_{x,y}^K\sum_{s=1}^{K-1} \omega^{s} R(N+1,s)\beta_y\Big|\\
&=\Big|\big\langle M', (R(x,s)\beta_y)_{x,s,y=1}^{N+1, K-1, N}\big\rangle\Big|.
\end{align*}}
By taking the supremum over all elements $(R(x,s)\beta_y)_{x,s,y}$, we prove the first inequality in \eqref{Estimates upper bound K}.
\subsection{Upper bounds in terms of the Hilbert space dimension}
\begin{proposition}\label{prop:boundsfordim}
For any asymmetric Bell inequality $M$ as in (\ref{eq:Bellineq}), if the quantum bound of $M$ is achieved by a quantum correlation in which Bob's local dimension is $d$, then
\begin{equation*}
B_{\mathcal{Q}}(M) \leq C \sqrt{d} B_{\mathcal{C}}(M),
\end{equation*}where $C$ is a universal constant.
\end{proposition}
The result can be proved by following the same ideas as in \cite[Theorem 3]{BV2013}. Here, we will provide a sketch of proof. Given $\epsilon>0$, let $|\Psi\rangle, (E_x^a)_{x,a}, (A_y)_y$ be a quantum state, some POVMs for Alice and dichotomic observables for Bob respectively, such that
\begin{equation*}
B_{\mathcal{Q}}(M) \leq  (1+\epsilon) \Big|\sum_{x,y,a} M_{x,y}^a \langle \Psi| E_x^a \otimes A_y |\Psi \rangle\Big|.
\end{equation*}
Note that $A_y \in B(H_d)$ for every $y.$ Let $|\Psi\rangle = \sum_{i=1}^d \lambda_i |u_i\rangle |v_i\rangle$ be the Schmidt decomposition, where $(|v_i\rangle)_i$ span the local Hilbert space of Bob. For every $y,$ let $M_y = \sum_{x,a}M_{x,y}^a E_x^a,$ and $E_{i,j} = \sum_y \langle v_i| A_y|v_j\rangle M_y$. Then, it is clear that $M=\sum_{i,j}E_{i,j}\otimes |v_i\rangle\langle v_j|$ verifies that $\big|\langle \Psi| M |\Psi \rangle \big|\geq (1+\epsilon)^{-1}B_{\mathcal{Q}}(M)$. On the other hand, since $M$ is hermitian we easily deduce that $E_{i,j} =E_{j,i}^{\dagger}$ for every $i,j$. Therefore, by the same argument as in \cite[Claim 7]{BV2013}, for every $i$ we have
\begin{equation*}
\max\Big\{\Big\| \sum_j E_{i,j} E_{i,j}^{\dagger} \Big\|,  \Big\| \sum_j E_{i,j}^{\dagger} E_{i,j}\Big\|\Big\} \leq 4 K_G^{\mathbb{C}} B_{\mathcal{C}}(M)^2.
\end{equation*}
The universal constant is slightly different from the one appearing in \cite[Claim 7]{BV2013}, since here we consider a bipartite case. Now, by the noncommutative Khinchine inequality \cite[Theorem 6]{BV2013}, one can show
\begin{align*}
\big|\langle \Psi| M |\Psi \rangle \big| \leq 6 (K_G^{\mathbb{C}})^{\frac{1}{2}}  \sqrt{d} B_{\mathcal{C}}(M).
\end{align*}
Since the argument is the same as in \cite[Theorem 3]{BV2013}, we omit the details.
\end{document}